\documentclass{icrc2009}

\usepackage{graphicx}   
\usepackage[caption=false]{caption}    
\usepackage[font=footnotesize]{subfig} 
\usepackage{fixltx2e}
\usepackage{url}

\newcommand{\shorttitle}[1]%
{\markboth{Proceedings of the 31\MakeLowercase{$^{st}$} ICRC, {\L}\'{o}d\'{z} 2009}{#1} }
\newcommand{\etal}{\MakeLowercase{\textit{et al. }}} 

\begin{document}
\title{Monitoring of Bright Blazars with MAGIC}

\author{\IEEEauthorblockN{
    Ching-Cheng Hsu\IEEEauthorrefmark{1},
    Konstancja Satalecka\IEEEauthorrefmark{2},
    Malwina Thom\IEEEauthorrefmark{3},
    Michael Backes\IEEEauthorrefmark{3},
    Elisa Bernardini\IEEEauthorrefmark{2}, 
                            \\
    Giacomo Bonnoli\IEEEauthorrefmark{4}, 
    Nicola Galante\IEEEauthorrefmark{7},
    Florian Goebel\IEEEauthorrefmark{1}\IEEEauthorrefmark{6},
    Elina Lindfors\IEEEauthorrefmark{5},
                            \\
    Pratik Majumdar\IEEEauthorrefmark{2},
    Antonio Stamerra\IEEEauthorrefmark{4} and
    Robert Wagner\IEEEauthorrefmark{1} on behalf of the MAGIC Collaboration}
                            \\
\IEEEauthorblockA{\IEEEauthorrefmark{1}Max-Planck-Institut fuer Physik, Foehringer Ring 6, D-80805 Muenchen, Germany.}
\IEEEauthorblockA{\IEEEauthorrefmark{2}DESY, Platanenallee 6, D-15738 Zeuthen, Germany.}
\IEEEauthorblockA{\IEEEauthorrefmark{3}Technische Universitaet Dortmund, D-44221 Dortmund, Germany.}
\IEEEauthorblockA{\IEEEauthorrefmark{4}Dipartimento di Fisica, Universit`a degli Studi di Siena, Via Roma 56, I-53100 Siena, Italy.}
\IEEEauthorblockA{\IEEEauthorrefmark{5}Tuorla Observatory, Dept. Physics and Astronomy, University of Turku, FI-20014 Turku, Finland.}
\IEEEauthorblockA{\IEEEauthorrefmark{7}Whipple Observatory, Harvard-Smithsonian Center for Astrophysics, P.O. Box 97, Amado, AZ 85645-0097, USA.}
\IEEEauthorblockA{\IEEEauthorrefmark{6}deceased}}
\shorttitle{Ching-Cheng Hsu \etal Monitoring of Bright Blazars }
\maketitle

\begin{abstract}
Blazars, a class of Active Galactic Nuclei (AGN) characterized by a close
orientation of their relativistic outflows (jets) towards the line of sight, 
are a well established extragalactic TeV $\gamma$-ray emitters. Since 
2006, three nearby and TeV bright blazars, Markarian (Mrk) 421, Mrk 501
and 1ES 1959+650, are regularly observed by the MAGIC telescope with single
exposures of 30 to 60 minutes. The sensitivity of
MAGIC allows to establish a flux level of 30\% of the Crab flux for each
such observation. In a case of Mrk 421 strong
flux variability in different time scales and a high correlation between
X-ray/TeV emissions have been observed. In addition, preliminary results
on measured light curves from Mrk 501 and 1ES1959+650 in 2007/8 
are shown.
\end{abstract}

\begin{IEEEkeywords}
Active Galactic Nuclei; BL Lacertae objects;
gamma-rays observations; gamma-ray telescopes
\end{IEEEkeywords}
 
\section{Introduction}
Blazars belong to the class of AGN and are characterized by
relativistic jets oriented towards the Earth. They have a continuous Spectral 
Energy Distribution (SED) with no or weak emission lines 
and two broad humps (one in the UV to soft X-ray and a second in the GeV-TeV
range). Moreover, their flux was found to be variable at all observed frequencies,
but on different time scales ranging form years to minutes \cite{Aharonian:2007ig} 
\cite{Albert:2007zd}. 

In recent years numerous multiwavelength campaigns were performed with the aim of
explaining the acceleration and emission mechanisms in blazars. In many campaigns
the new generation of Imaging Atmospheric Cerenkov Telescopes (IACTs)
like HESS \cite{Aharonian:2006pe}, MAGIC \cite{Albert:2007xh} and VERITAS 
\cite{Acciari:2008ah} also took part allowing us to have a deeper look at the
highly variable VHE (E $\ge$ 100\,GeV) $\gamma$-ray emission. Unfortunately the data collected 
so far is not yet enough to fully constrain the theoretical models and still one of the 
most important question remains unanswered: 
are the leptonic or hadronic acceleration processes responsible for the observed blazar 
behavior? 

Leptonic models, like for example, the Synchrotron-Self Compton (SSC) \cite{Tavecchio:1998xw} 
are very successful in describing most of the existing SEDs and offer a reasonable 
explanation for the fast variability of blazars. Hadronic models, on the other hand, 
like the Synchrotron Mirror Model (SMM) \cite{Reimer:2005sj} or Synchrotron Proton
Blazar (SPB) \cite{Muecke:2002bi} apart from a good description of the SED
structure can also explain the ``orphan'' $\gamma$-ray flares (see e.g. \cite{Krawczynski:2003fq}) 
and predict emission of high energy neutrinos.

\section{AGN monitoring}
As mentioned before the new generation IACTs can give a valuable input for 
understanding of the acceleration mechanism in blazars. Not only by participation 
in multiwavelength observations, but also by
performing a source state independent, long term monitoring of the most interesting brighter
$\gamma$-ray emitters.  
There are many advantages of such observations. They allow to obtain an unbiased distribution 
of flux states and perform any statistical study which requires high statistics on 
various flux levels. 
For example: the determination of flaring state probabilities, essential for the estimation 
of the statistical significance of possible correlations between flaring states
and other observables, such as neutrino events \cite{Satalecka:2007}. In view of the 
results expected from the IceCube neutrino observatory \cite{Ahrens:2003ix} such a study
is of particular interest.

Investigation of spectral changes occurring during periods of different source activity may 
also allow to improve our knowledge about the acceleration and emission processes.

Another important aspect of the AGN monitoring is triggering Target of Opportunity
(ToO) observations. These follow-up observations may be performed by the IACT issuing
the ToO trigger, include other IACTs -- allowing to increase the time coverage of the 
observations -- or telescopes and satellites observing at other wavelengths. 
In a context of ``orphan'' TeV flares simultaneous X-ray observations are especially 
valuable.

\section{The MAGIC Telescope}

MAGIC is currently the largest single-dish IACT for VHE $\gamma$-ray astronomy. It is located on
the Canary Island of La Palma, at an altitude of 2200\,m a.s.l and has been
in scientific operation since summer 2004. 
In January 2007 a major upgrade of the MAGIC Telescope with
a Multiplexed Fiber-Optic 2\,GSamples/s FADC Data Acquisition system took place
\cite{Goebel:2007cb}. The fast readout minimizes the influence of the background
from the light of the night sky and the additional information on the time structure 
of the shower signal helps to reduce the hadronic background \cite{Aliu:2008pd}. 
The trigger threshold of MAGIC is around 60\,GeV\footnote{The trigger threshold is 
defined as the peak of the energy distribution of the triggered events.}.
A source emitting $\gamma$-rays at a flux level of 1.6\% of the Crab Nebula can be detected with 5\,$\sigma$
significance within 50 hours of observation time. This sensitivity is sufficient
to establish a flux level of about 30\% of the Crab flux above 300\,GeV for a
30\,min observation (Fig.1). A quick on-line analysis system allows to estimate
the flux level of the observed source and send ToO triggers on high flux levels
during data taking.
At present MAGIC is the only IACT which can observe under moderate moon and twilight 
conditions with only slightly lower sensitivity.
The construction of a second telescope (MAGIC-II) is being finalized
and it is planned to start stereoscopic observations in autumn 2009 \cite{Cortina:icrc2009}.
\begin{figure}[!t]
 \centering
  \includegraphics[height=2.3 in]{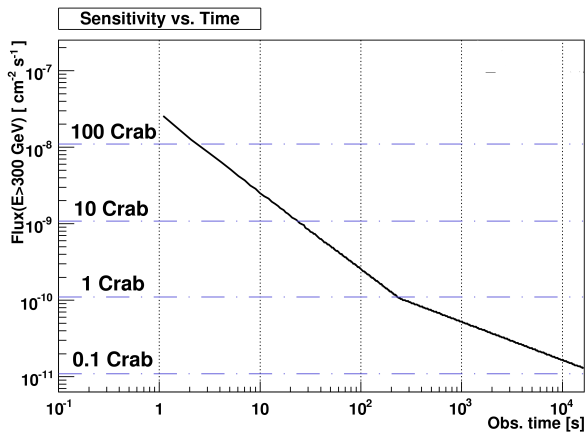}
  \caption{MAGIC sensitivity corresponding to a 5\,$\sigma$ detection \cite{LiMa:1983} as a function of exposure time.}
\end{figure}

\section{Monitoring strategy}

In order to achieve a dense sampling, up to 40 short observations per source
are scheduled, evenly distributed over the observable time by MAGIC. 
Three sources were chosen for a regular monitoring: Mrk 421, Mrk 501
and 1ES 1959+650. The first two are relatively bright and usually 15-30\,min
observations are scheduled for them. 1ES~1959+650 being fainter requires
longer observation times, at least 30 minutes per single exposure.

The major part of the monitoring ($\sim$60\%) has been performed under moderate moonlight or twilight, 
keeping the impact on the overall observation schedule low and allowing to maximize the
available duty cycle up to $\sim$12\%.

\section{Results}

In this section we present the preliminary results of the MAGIC AGN
monitoring program for the observation season 2007/2008 together with some previously 
taken, published MAGIC data on the same objects. The data have been
processed with the standard MAGIC analysis tools \cite{Albert:2007xh}.
A fraction of the data has been removed due to poor observation conditions.
All cuts were optimized and verified with Crab Nebula data.

\subsection{Mrk 421}
Between February 2007 and June 2008 82 hours of data from Mrk 421 were taken. 
The observations were mostly performed in wobble mode, which allows to simultaneously 
collect signal and background events. About 66 hours of good quality data
(80\%) were used for further analysis. It should be noted that about 70\% of these data were 
taken due to the ongoing flaring activity of the source and are actually not 
part of the monitoring campaign.

\begin{figure*}[th]
  \centering  
  \includegraphics[width=5in]{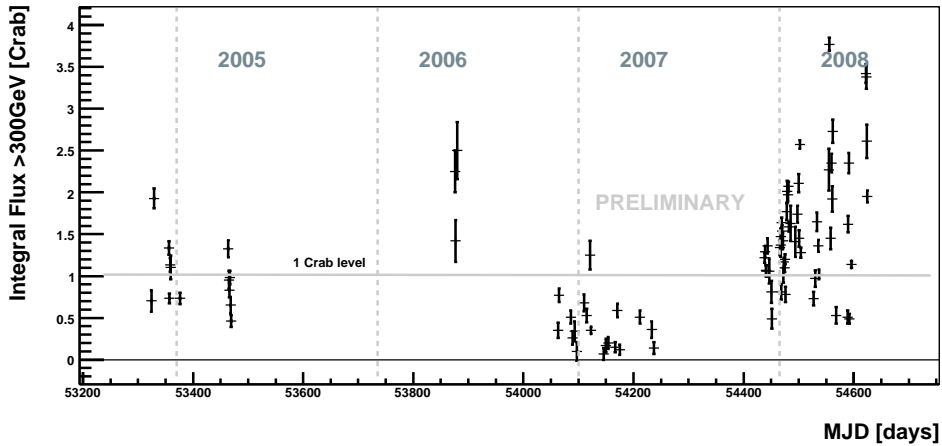}
  \caption{\label{LC_Mrk421_all}
Mrk 421 light curve showing all data collected by MAGIC \cite{Albert:2006jd} \cite{Goebel:2007uu} so far.
}
\end{figure*}

All observations of Mrk 421, performed by MAGIC since 2004 \cite{Albert:2006jd}
\cite{Goebel:2007uu}, are shown in Fig.\ref{LC_Mrk421_all}. The source was very 
active in 2008: many flares were observed and flux rarely decreased below 1 Crab level
(F$_{E>300\,GeV}$ = $1.23\pm 0.10)\times10^{-10}$\,ph/cm$^{-2}$s$^{-1}$ \cite{Albert:2007xh}).   
According to a ToO agreement with HESS and VERITAS, MAGIC issued several alerts during
this time of high activity. 
A detailed analysis of the collected data, such as a study on the intra-night variability 
is discussed in \cite{Bonnoli:icrc09}.

As mentioned before, possible correlations between different
wavelengths, in particular TeV $\gamma$-rays and X-rays, are an
interesting subject to investigate. We therefore searched for X-ray 
and optical measurements which are within 6\,h before or after the TeV observations.
In this work we used the X-ray data taken by the ASM instrument aboard of the RXTE satellite, 
which are publicly available on the project\footnote{http://xte.mit.edu/ASM\_lc.html} web page. 
The averages and errors of ASM data points were calculated
on a dwell-by-dwell (90 seconds) data basis.
If the number of dwells were fewer than five, we discarded that data point.
We finally selected 75 pairs of TeV-X-ray measurements for further analysis.

The 1.03\,m telescope at the Tuorla Observatory Finland and the 35\,cm KVA telescope 
at La Palma, Canary Islands provided us with the optical $R$-band data from their
Blazar Monitoring Program\footnote{http://users.utu.fi/kani/1m/}. 
The optical flux was corrected for the flux of the host galaxy and the flux contribution 
from the companion galaxies \cite{Nilsson:2007ax}. In the end, 56 measurements pairs were 
found.

\begin{figure*}[t]
   \centerline{\subfloat{\includegraphics[width=2.5in]{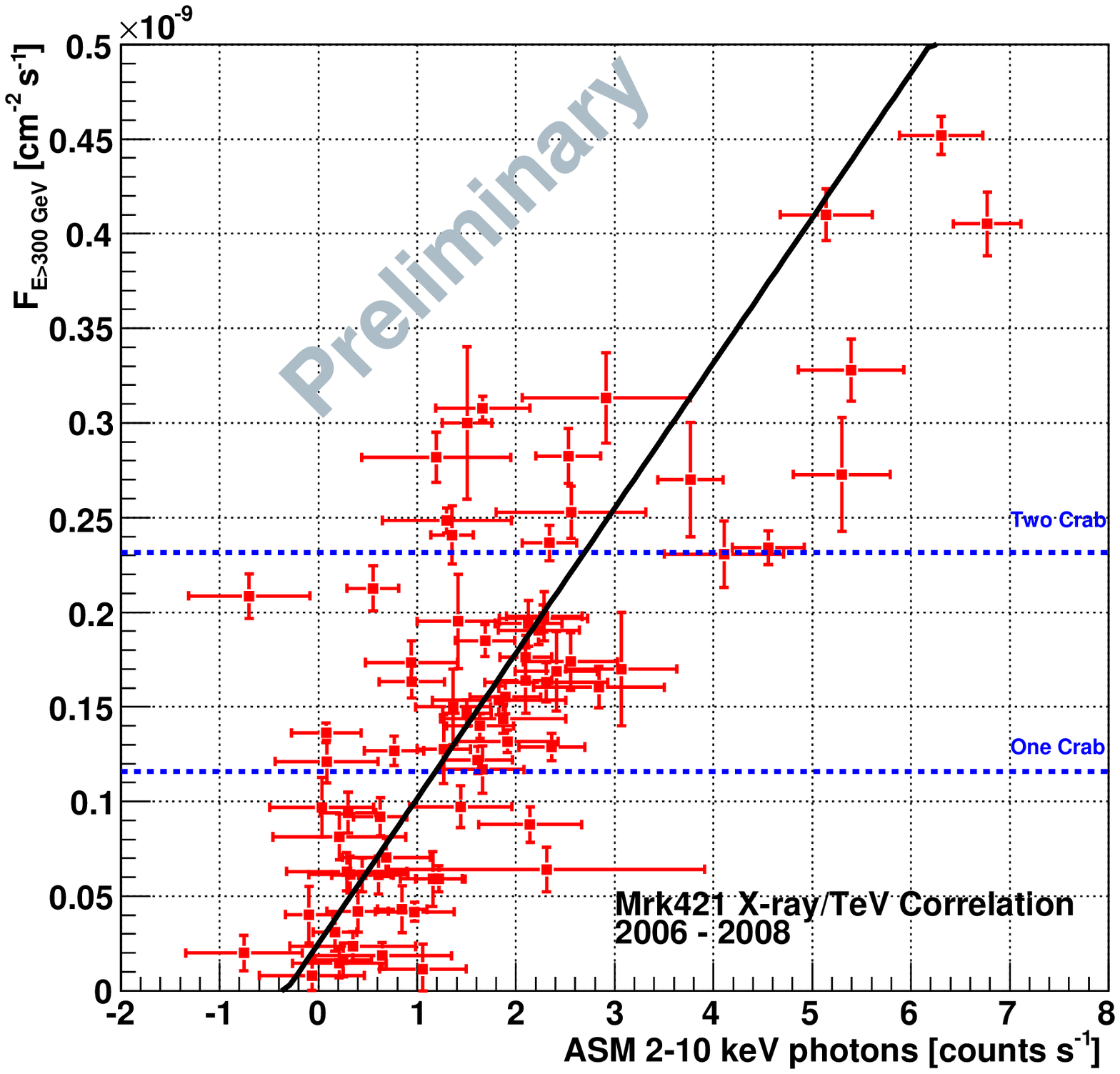} \label{TeVXray}}
              \hfil
              \subfloat{\includegraphics[width=2.5in]{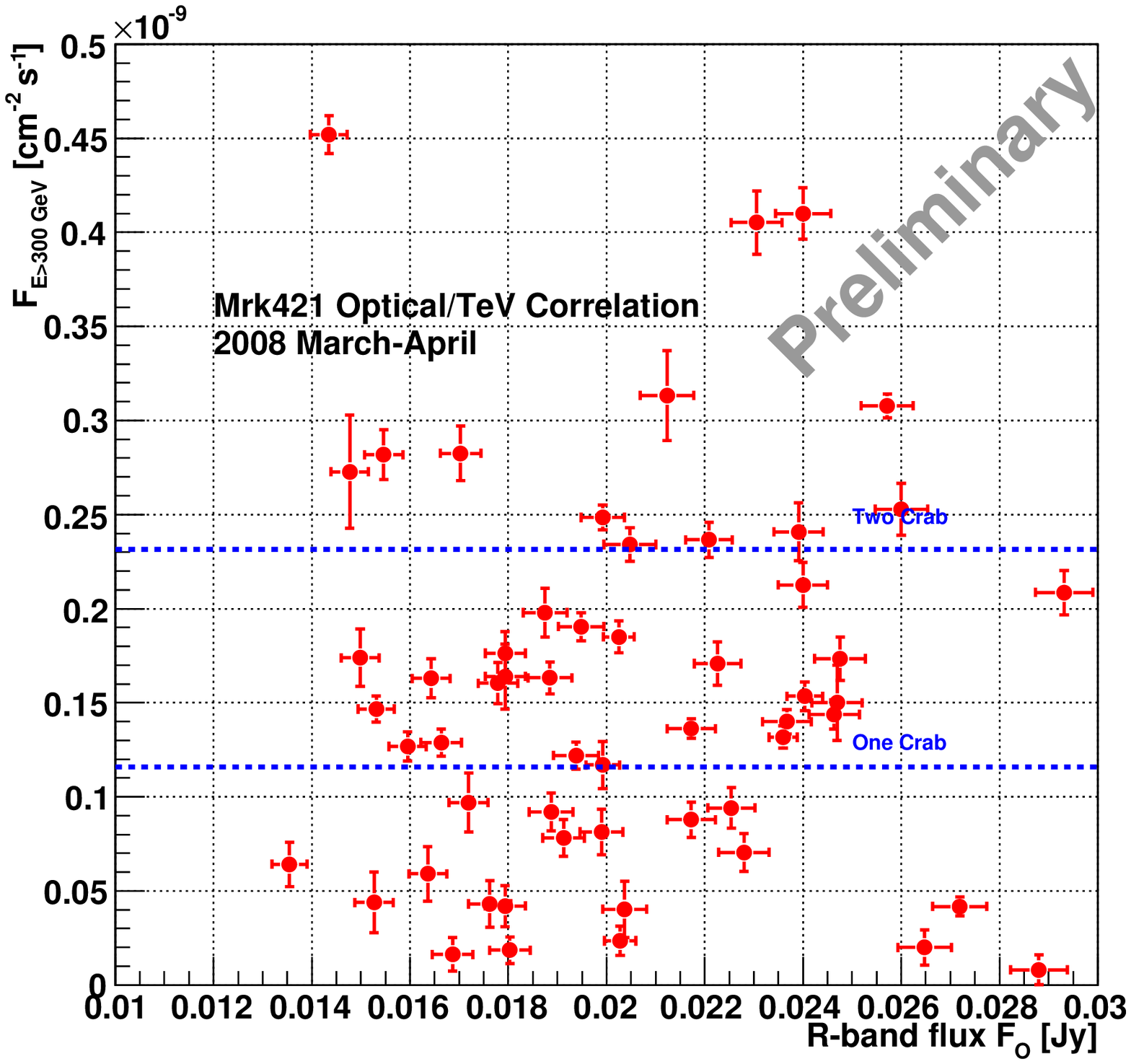} \label{TeVoptical}}
             }
   \caption{Left: VHE $\gamma$-ray (MAGIC) - X-ray (ASM) correlation plot for Mrk 421.
           Right: VHE $\gamma$-ray (MAGIC) - $R$-band (KVA) correlation plot for Mrk 421.}
   \label{double_fig}

 \end{figure*}

We used the Pearson's method (see section 14.5 in \cite{638765}) to calculate the correlation 
coefficients. For the X-ray/TeV data set (shown in Fig.\ref{double_fig} left), the correlation coefficient 
value is r~ = 0.77$\pm$0.05 and for optical/TeV data (Fig.\ref{double_fig} right), the coefficient 
becomes r = 0.03$\pm$0.14.

The significant (8\,$\sigma$) correlation with X-rays we found might point to a leptonic origin of the 
emission (see e.g. \cite{Wagner:2008cw} and references therein) but certain hadronic 
models also predict such a correlation (see e.g. \cite{Muecke:2002bi} or \cite{Aharonian:2000pv}).
In any case, conclusions on the origin or mechanism of the emission 
should be made very carefully because the data were not taken strictly simultaneously. 
%

\subsection{Mrk 501}

\begin{figure*}[th]
   \centering 
  \includegraphics[width=5in]{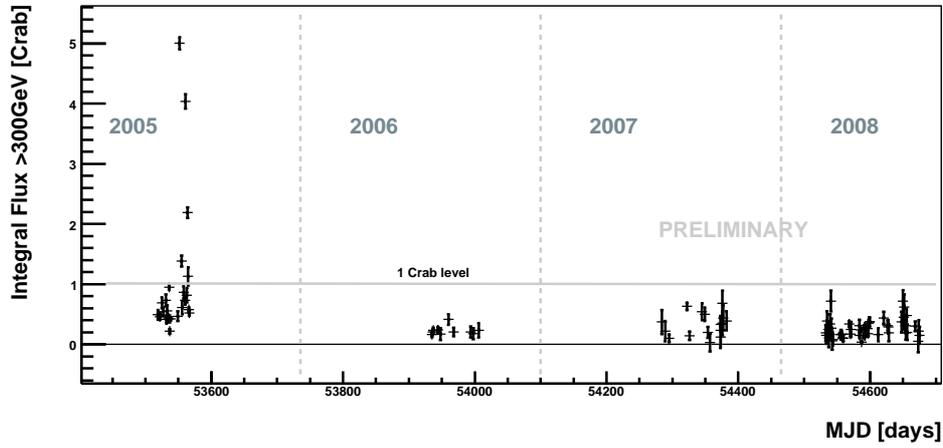}
  \caption{\label{LC_Mrk501_all}
Mrk 501 light curve showing all data collected by MAGIC \cite{Goebel:2007uu} \cite{Albert:2007zd} so far.
}
\end{figure*}

Fig. \ref{LC_Mrk501_all} shows all the data collected from Mrk 501 by the MAGIC
telescope since 2005, when the source was found in a flaring state and doubling
times as short as few minutes were observed \cite{Albert:2007zd}.
The new results presented here are based on the data collected between February
2007 and August 2008. As in the case of the Mrk 421, the observations were performed 
in wobble mode, mostly
during moderate moonlight or twilight in order to maximize the time coverage
of this source (56\% of the total observation time). After quality selection
16 hours of data remained and were analyzed.
A part of the light curve presented in Fig. \ref{LC_Mrk501_all} 
was taken during a multiwavelength campaign (MJD 54550-54602) described in more 
detail in \cite{Kranich:icrc09}.
Similar to the year 2006 \cite{Goebel:2007uu} in the 2007/2008 observational period
Mrk 501 was in a relatively low state (below 1\,Crab). Thanks to good weather conditions 
at the site a dense sampling was obtained. A statistical analysis, e.g. estimation of the 
source state probability, is in progress.

\subsection{1ES 1959+650}
\begin{figure}[t]
  \centering
  \includegraphics[width=2.5in]{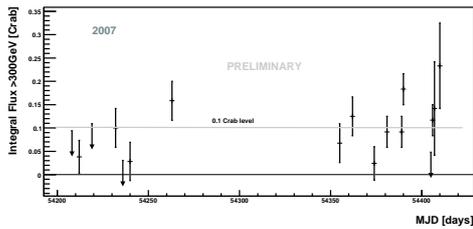}
  \caption{\label{LC_1ES1915_07}
1ES 1959+650 light curve showing MAGIC data from the year 2007. The arrows denote upper limits
at 90\% C.L.
}
\end{figure}
MAGIC monitored 1ES 1959+650 from April 2007 till October 2008 for 
27 hours under large zenith angle (35$^{\circ}$-50$^{\circ}$) conditions.
The source was observed in wobble mode.
After quality selection mainly based on bad atmospheric conditions and unstable, low event 
rates, 12.7 hours of data taken in 2007 and 3.6 hours of data taken in 2008 were analyzed.

In 2007 a clear $\gamma$-ray signal of 9.0\,$\sigma$ is seen.
However, the mean 1ES1959+650 $\gamma$-ray flux above 300\,GeV during this observations 
period with MAGIC was only $(0.92\pm 0.12)\times10^{-11}$\,ph/cm$^{-2}$s$^{-1}$, i.e. 
$\sim$10\% Crab.
The overall light curve for 1ES1959+650 in 2007 is shown in Fig. \ref{LC_1ES1915_07},
all measurements with significances below 1\,$\sigma$ were converted to flux upper 
limits. The light curve indicates no major changes of the flux level and no significant 
flares. The overall significance of the data sample from 2008 is 2.6 $\sigma$, which allows us 
to set an upper limit on the flux above 300\,GeV of 1.54$\times10^{-11}$\,ph/cm$^{-2}$s$^{-1}$ 
at 90\% C.L.
In comparison to previous observations \cite{Aharonian:2003} 
\cite{Gutierrez:2006ak} it can be concluded that MAGIC observed 1ES1959+650 in 2007/2008 during 
its usual quiescent state.

\section{Conclusions}

During the observational season 2007/2008 three blazars were regularly monitored by MAGIC:
Mrk 421, Mrk 501 and 1ES 1959+650. Here we presented preliminary results of the measured
flux levels for all three sources. Mrk 501 and 1ES 1959+650 were 
found in a low state, but the dense sampling of Mrk 501 provides valuable material 
for further statistical studies. Mrk 421 has shown an interesting flaring activity in 2008.
We also investigated possible correlations of the TeV and X-ray/optical flux levels 
for Mrk 421. We found a significant correlation between TeV and X-rays but no correlation 
with the optical $R$-band. Parts of the data collected from Mrk 421 and Mrk 501 are discussed 
in more detail in dedicated contributions \cite{Bonnoli:icrc09} \cite{Kranich:icrc09}.

\section{The Acknowledgments}
We would like to thank the Instituto de Astrofisica de Canarias for the
excellent working conditions at the Observatorio del Roque de los Muchachos
in La Palma. The support of the German BMBF and MPG, the Italian INFN and
Spanish MCINN is gratefully acknowledged. This work was also supported by
ETH Research Grant TH 34/043, by the Polish MNiSzW Grant N N203 390834,
and by the YIP of the Helmholtz Gemeinschaft.

\end{document}